# Magnetothermopower and magnetoresistance of single Co-Ni/Cu multilayered nanowires

Tim Böhnert[1], Anna Corinna Niemann[1], Ann-Kathrin Michel[1], Svenja Bäßler[1], Johannes Gooth[1], Bence G. Tóth[2], Katalin Neuróhr[2], László Péter[2], Imre Bakonyi[2], Victor Vega[3], Victor M. Prida[3], and Kornelius Nielsch[1]

1. Institute of Applied Physics, Universität Hamburg, Hamburg, Germany.
2. Wigner Research Centre for Physics, Hungarian Academy of Sciences, Budapest, Hungary.
3. Depto. Física, Universidad de Oviedo, Oviedo, Spain.

The magnetothermopower and the magnetoresistance of single Co-Ni/Cu multilayered nanowires with various thicknesses of the Cu spacer are investigated. Both kinds of measurement have been performed as a function of temperature (50 K to 325 K) and under applied magnetic fields perpendicular to the nanowire axis, with magnitudes up to -15 % at room temperature. A linear relation between thermopower $S$ and electrical conductivity $\sigma$ of the nanowires is found, with the magnetic field as an implicit variable. Combining the linear behavior of the $S$ vs. $\sigma$ and the Mott formula, the energy derivative of the resistivity has been determined. In order to extract the true nanowire materials parameters from the measured thermopower, a simple model based on the Mott formula is employed to distinguish the individual thermopower contributions of the sample. By assuming that the non-diffusive thermopower contributions of the nanowire can be neglected, it was found that the magnetic field induced changes of thermopower and resistivity are equivalent. The emphasis in the present paper is on the comparison of the magnetoresistance and magnetothermopower results and it is found that the same correlation is valid between the two sets of data for all samples, irrespective of the relative importance of the giant magnetoresistance or anisotropic magnetoresistance contributions in the various individual nanowires.

# I. INTRODUCTION

The thermopower, or Seebeck coefficient, $S$ describes the redistribution of charge carriers driven by an applied temperature gradient. Experiments show a linear behavior between thermopower $S$ and electrical conductivity $\sigma$[1–14], with the magnetic field as an implicit variable, in metals with anisotropic magnetoresistance (AMR)[15–17] or giant magnetoresistance (GMR)[18,19] effects. Comparing this linear relation to the Mott formula[20], which describes the diffusive part of the thermopower, a direct proportionality between $S$ and $\sigma$ is predicted in materials with negligible nondiffusive contributions. On the contrary, the experimental results do not obey these clear predictions and often a more complicated relationship is presumed. The major experimental difficulty is that only relative Seebeck coefficients of the specimen with respect to the contact material are accessible. In order to obtain absolute thermopower values, the measured data have to be corrected by the absolute values of the contact material. These absolute literature values are calculated ultimately from observations of the Thomson effect of pure bulk samples[21,22]. Since the thermopower is very sensitive to impurities[23] and shows size effects[24,25], deviations between the literature values and properties of materials used in the experiments should be considered. However, an evaluation of a material's absolute Seebeck coefficient directly from thermopower measurements has not been challenged to date. In this work, the different thermopower contributions are distinguished utilizing a simple model based on the Mott formula. This enables us to separate the different thermopower contributions of the specimen and the electrical contact structure.

Currently, the interest in the magnetothermopower (MTP) of magnetic nanostructures is high, as recent publications reporting on measurements on single nanowires[26,27], tunnel junctions[28], and spin valves[29,30] show. Especially multilayered nanowires constitute a perfect model system to experimentally investigate spin-dependent transport in the current-perpendicular-to-plane (CPP) mode. The CPP transport is of particular interest in the concept of spin heat accumulation, which is proposed to cause a violation of the Wiedemann-Franz law[31].

According to the literature, electrodeposited Co-Ni/Cu nanowires[32–36] exhibit higher GMR values (between −23% and −35%) than Co/Cu nanowires[1,37–44] (between −14% and −15%). The magnetic-field dependence of $S$ in materials that show AMR or GMR effects is explained either by two spin-dependent Seebeck coefficients[1,2] or through the Mott formula using the resistivity instead of the electrical conductivity of the sample[1–9]. The latter approach has been applied by Conover *et al.*[3] and is used to predict equivalent MTP and magnetoresistance (MR) be-havior. However, in spite of several related publications, this prediction has not been convincingly demonstrated[8,27,45]. One of the few publications on this topic is by Costache *et al.*[46], who managed to separate the magnon contribution from the diffusive thermopower of ferromagnetic thin films. In addition, the energy derivative of the resistivity from the Mott formula can be calculated, which can be correlated with the transmission function serving as a starting point in theoretical models.

To contribute to a deeper insight into the interplay between heat and spin, Co-Ni/Cu multilayered nanowires are electrochemically deposited into nanoporous alumina templates. Electrical contacts are lithographically defined on top of single nanowires on a glass substrate in contrast to measurement approaches performed on platforms[47–50], in which the particular nanowire had to be assembled on top of a predefined structure. The thermopower and magnetoresistance of several single nanowires have been measured in a wide temperature 1098-0121/2014/90(16)/165416(11) 165416-1 ©2014 American Physical Society TIM BÖHNERT *et al.* PHYSICAL REVIEW B **90**, 165416 (2014) range as a function of the magnetic field. The relation between these two properties is compared to the Mott formula in order to extract the absolute thermopower of the nanowire materials. The paper is organized as follows. Section II gives a definition and the theoretical background for the main experimental quantities (MR and MTP) investigated in the present paper. In Sec. III, the nanowire synthesis and the measurement setup is explained in detail. The results of MR and MTP measurements are presented in Sec. IV, while Sec. V is devoted to a discussion of their correlation based on the Mott formula. Finally, the conclusions are summarized in Sec. VI.

## II. THEORETICAL BACKGROUND

### A. Magnetoresistance

The crucial parameters for an understanding of the magnetotransport phenomena in nanowires are the resistivity $\rho$ and the magnetoresistance[15–17]. In the following, two definitions of the MR ratio $r_{MR}$ are compared to the MTP ratio $r_{MTP}$, namely, the conservative $r_{MR} = (\rho_H - \rho_0)/\rho_0$ and the inflationary (also called optimistic) $r_{MR,inf} = (\rho_H - \rho_0)/\rho_H$, with the zero-magnetic-field resistivity $\rho_0$ and the resistivity in the magnetic field $\rho_H$. In multilayered nanowires, which are the subject of the present study, an appropriate alternating sequence of magnetic and nonmagnetic segments leads to spin-dependent scattering events that result in a GMR effect[18,19] and simultaneously the spin-dependent scattering events within the magnetic layer lead to an AMR effect[15–17]. The relative importance of the GMR and AMR effects in a given nanowire depends on the individual layer thicknesses, interfacial features, and the eventual presence of pinholes in the nonmagnetic spacer layer.

## B. Thermopower

The origin of the thermopower or Seebeck coefficient lies in the temperature-dependent average energy of the electrons, which generally leads to a diffusion towards the cold side. The accumulation of charge carriers at the cold side builds up the so-called thermoelectric voltage. The proportionality factor between the thermoelectric voltage $U_{thermo}$ and the temperature difference $\Delta T$ across a sample is called the Seebeck coefficient $S$. This coefficient is an intrinsic material property and is defined either absolutely against a superconductor or relative to a specific metallic material (mostly platinum). Per definition, the Seebeck coefficient is negative if electrons diffuse towards the cold side of the sample. The Mott formula is a firstorder approximation of the Boltzmann transport equation and describes $S$ in the free-electron model[20]. Substituting the electrical conductivity for the resistivity, the following expression for the Mott formula is derived:

$$S = -\frac{cT}{\rho}\left(\frac{d\rho(E)}{dE}\right)_{E=E_F}, \qquad (1)$$

with $c = \pi^2 k_B^2/3q$, $q$ being the charge of the carriers, $k_B$ the Boltzmann constant, $\rho$ the electrical resistivity, and $E$ the energy of the charge carriers. The energy derivatives of the resistivity are attainable by first-principle calculations, but only a fewpublications consider the band structure[10,51].

The thermopower expressed by the Mott formula describes solely diffusive contributions of the thermopower, which is valid only if the charge carriers are scattered dominantly by impurities and lattice defects. However, at certain temperatures, electron-phonon and electron-magnon collisions can give rise to additional thermopower contributions called phonon drag and magnon drag, respectively[52–54]. The diffusive thermopower is independent of the phonon drag, as it does not change the heat capacity of the electrons[23,52]. Therefore, the diffusive and nondiffusive thermopower contributions are independent of each other and simply add up.

The relative change of the Seebeck coefficient in an applied magnetic field is called magnetothermoelectric power (MTEP) ratio and is defined as:

$$r_{MTEP} = \frac{U_{thermo}(H) - U_{thermo}(0)}{U_{thermo}(0)}. \qquad (2)$$

Here $U_{thermo}$ describes the measured thermoelectric voltage with respect to the contact material. In the case of opposite signs of the thermopower of the nanowire and the contact material, the

$r_{MTEP}$ can reach infinite values and should be treated with caution. The term magnetothermopower ratio will be used to describe the magnetic-field-induced effect relative to the absolute Seebeck coefficient of the nanowire sample and is defined as:

$$r_{MTP} = \frac{S_{NW}^{abs}(H) - S_{NW}^{abs}(0)}{S_{NW}^{abs}(0)} = \frac{U_{thermo}(H) - U_{thermo}(0)}{U_{thermo}(0) - S_{contact}^{abs}(0) \cdot \Delta T} \quad (3)$$

with the absolute thermopower values $S^{abs}$. Obtaining the $r_{MTP}$ is experimentally challenging due to uncertainties of the absolute Seebeck coefficient value of the contact materials *S*abs contact. Absolute literature values for most metals from low temperature to room temperature (RT) are available[55], but the Seebeck coefficient is very sensitive to impurities[23] and size effects[24,25]. Therefore, it depends on the fabrication technique and deviations between the literature values and properties of materials used in the experiments should be considered.

The magnetic field dependence of the energy derivatives of the resistivity in the Mott formula and the relation between $r_{MTP}$ and $r_{MR}$ are the focus of the present work. One of the few publications on this topic by Conover *et al.*[3] directly relates $r_{MTP}$ and $r_{MR}$ by utilizing the linear relationship between $S$ and $\sigma$ discovered by Nordheim and Gorter[56,57]. Up to now, in several magnetic systems, such as granular alloys, magnetic/nonmagnetic multilayers, spin valve structures, and alloys, a linear dependence of the Seebeck coefficient on the electrical conductivity in an applied magnetic field has been found[1–14]. By comparing the linear relationship to Eq. (1), it seems reasonable to assume that the quantity $d\rho/dE$ at the Fermi energy does not depend on the magnetic field.

## III. EXPERIMENT

### A. Sample preparation and characterization

In this work, self-ordered anodized aluminum oxide membranes exhibiting a hexagonal nanoporous structure[58] were used as templates for the preparation of Co-Ni/Cu nanowires. In order to passivate the surface of the nanowires, the pores of the alumina template were coated by atomic layer deposition with a $SiO_2$ shell having a thickness of about 5 nm[59,60]. The multilayered Co-Ni/Cu nanowires were prepared by two-pulse plating from a single bath by using an Ivium CompactStat potentiostat. The Cu layers were electrodeposited at a potential of −0.58 V with respect to an Ag/AgCl electrode, while the Co-Ni layers were electrodeposited at a potential of −1.5 V vs an Ag/AgCl electrode. A $Co_{0.5}Ni_{0.5}$ composition was chosen to maximize the GMR effect[61] according to the recipe published by Tóth *et al.*[62]. The applied Cu deposition potential corresponds to the electrochemically optimized value[62] at which neither a dissolution of the magnetic layer during the Cu deposition nor a codeposition of magnetic atoms into the Cu spacer layer occurs. A Cu content of about 2% is estimated in the magnetic layers based on the ratio of the current densities during the two pulses[62]. According to the results of Pullini and Busquets-Mataix on multilayered nanowires[63], the current efficiencies can be expected slightly below the value of 100% known from depositions on thin films[62]. Furthermore, in agreement with previous reports[63], there is some uncertainty of the effective cathode area due to a canting of layer planes with respect to the nanowire axis up to 45° [see Figs. 1(a) and 1(b)] and due to an unknown amount of filled pores. Thus, the nominal deposition values have significant unsystematic errors and detailed electron microscopy investigations are conducted to get a reliable estimate of the actual layer thicknesses by using the bilayer thickness and the analyzed nanowire composition. The multilayered nanowire samples will be identified by the Cu layer thickness obtained from these studies in a manner as described below.

After electrodeposition of the nanowires, a mixture of chromic acid and phosphoric acid was used to dissolve the template selectively. The single nanowires with diameters around 240 nm were separated by filtration from the solution, diluted in ethanol, and finally deposited on a glass substrate with a thickness of 150 $\mu$m. Transmission electronmicroscope micrographs of different nanowire sections of the 3.5- and 5.2-nm Cu samples are shown in Fig. 1. A chemical analysis of the 0.9-, 1.4-, 3.5-, and 5.2-nm Cu samples was performed by transmission electron microscopy energy dispersive x-ray spectroscopy (TEMEDX) (JEM 2100) on single nanowires placed on a fine-mesh copper grid. For the 3.5-nm Cu sample, the composition was also measured by scanning electron microscopy energy dispersive x-ray spectroscopy (ZEISS SIGMA) on the cross section of the membrane. The overall composition of the Co-Ni/Cu nanowires is given in columns 4 and 5 of Table I. The uncertainty of the composition analysis is around 5% for each element and the statistical deviation between individual nanowires of the same sample is in the same range. The homogeneity of the composition along and across the nanowire axiswas confirmed by TEMEDX. The TEM analysis results were corrected for the average Cu background signal due to the transmission electron microscope copper grid, which leads to an additional error for the TEMEDX results. The large scatter of the Co-Ni composition in the magnetic layers of different samples, exceeding themeasurement error, may be from the preparation conditions. The composition of the magnetic layer is known to change very rapidly with the ionic concentration ratio in the template and the deposition current density[64]. These factors are difficult to control due to complex diffusion conditions in the template and might contribute to the differences observed in the magnetic layer compositions of the various samples.

In addition to the quantitative chemical analysis, the bilayer thickness was determined from TEMEDX line scans along the nanowires at several positions and averaged. For the 3.5-nm Cu sample, the bilayer length was also estimated from scanning electron microscope micrographs by dividing the average nanowire length by the number of bilayer pulses during electrodeposition. With information about the bilayer thickness and the overall nanowire composition, the average thickness of the magnetic and nonmagnetic layers was determined with an uncertainty of about of the layer thicknesses, the amount of Cu in the magnetic layer is insignificant and was neglected. Nevertheless, these are average values only and do not necessarily agree with the actual data of the individual nanowire used in the magnetotransport measurements. Despite all the uncertainties, the SEM and TEM results are in good agreement (as shown in Table I for the 3.5-nm Cu sample), which gives a hint at the reliability of the analysis. The determination of the layer thicknesses of electrodeposited multilayered nanowires is still a remaining problem[32].

It should be kept in mind that the uncertainties of the chemical composition as well as the bilayer length and the individual layer thicknesses do not have a significant impact on the analysis of the measured magnetotransport data. This is because the main emphasis in the present paper is on a comparison of the magnetoresistance and magnetothermopower results measured on the same individual nanowire. As will be presented in Sec. IV, the same correlation was found to be valid between the two sets of data for all samples, irrespective of the relative importance of the GMR or AMR contributions, which is finally determined by the actual thicknesses of the magnetic and nonmagnetic layers. Therefore, the geometrical and compositional properties are rather used to label the samples.

For the determination of the resistivity, knowledge of the diameter and length of the nanowire is necessary. These are determined from scanning electron microscope micrographs of each nanowire and corrected for the passivating $SiO_2$ layer thickness determined from transmission electron microscope images such as those shown in Fig. 1. Figure 1(d) shows part of a nanowire with a damaged $SiO_2$ layer. The upper part of the Cu layers is dissolved, which leads to the high contrast of the micrograph and is not representative for all samples.

For the present work, six multilayered nanowire samples were altogether prepared with different deposition parameters, which were grouped according to their measured magnetoresistance characteristics as shown in Fig. 2. The 3.5-nm Cu sample was designed to have approximately equal magnetic and nonmagnetic layer thicknesses and exhibited the largest GMR effect with a very small AMR contribution. The other five samples were designed to have various Cu layer thicknesses and a constant magnetic layer thickness. The 0.8-, 1.4-, and 5.2-nm

Cu samples exhibit significant GMR and AMR contributions, while the 0.2- and 0.9-nm Cu samples show solely the AMR effect.

## B. Magnetotransport measurements

In the present work, the magnetotransport has been measured on single nanowires with the help of lithographically defined measurement platforms on a glass substrate. For this purpose, a double layer of photoresist (ma-P 1205, microresist technology) and lift-off resist (LOR-3B, Micro Chem)was applied. Using a laserwriter ($\mu$PG 101, Heidelberg Instruments), the photoresist was exposed and developed. An *in situ* radio-frequency argon sputter etching was applied for 15 min to remove the SiO2 shell and surface oxides of the nanowire in order to achieve low-resistance Ohmic contacts. The electrical contacts consisted of a few-nanometer-thick titanium adhesion layer and a 60-nm-thick platinum layer, both sputtered prior to the lift-off process. A scanning electron microscope micrograph of a typical electrical contact structure for thermoelectric characterization of a single nanowire is shown in Fig. 3(a). The microheater was located perpendicular to the nanowire to create a temperature gradient of 3 K $\mu$m$^{-1}$ via Joule heating. Two resistance thermometers were placed at a distance of 8 $\mu$m along the nanowire and served three purposes: probing the temperature difference $\Delta T$ as well as measuring both the resulting thermovoltage $U$thermo between the hot and the cold contacts to the nanowire and the electrical conductivity $\sigma$. This electrical contacting method has been described elsewhere[9].

With the setup introduced above, temperature-dependent measurements could be performed with the magnetic field perpendicular to the nanowire axis in a cryostat. During a full measurement cycle, the temperature of the cryostat $T_{\text{cryostat}}$ was changed stepwise between 50 and 325 K. At each increment, the magnetic field was varied stepwise up to ±3 T and electric powers of roughly 1, 2, and 5 mW were applied to the heater. The resistance or the thermovoltage of the nanowire was measured separately at thermal equilibrium, which is necessary to reduce the noise level. All resistance values were measured with the four-point technique at 128 and 189 Hz by the lock-in amplifier of the cryostat with a feedback-controlled alternating current source at 10 $\mu$A. The thermovoltage was measured by a 2182A nanovoltmeter (Keithley) with an input impedance greater than 10 G$\Omega$ and corrected by the offset at zero heat voltage (about 1 $\mu$V). To probe the temperature difference, the ac resistance of both thermometers was measured and calibrated against $T$cryostat at zero heat power. The temperature difference $\Delta T = T_{\text{hot}} - T_{\text{cold}}$ is about one-third of the rise of the average temperature of the nanowire $\overline{T} = (T_{\text{HOT}} + T_{\text{COLD}})/2$ for the electrical contact structure shown in Fig. 3(a). A so-called probe-station setup was used for measurements at RT with the magnetic field perpendicular and parallel to the nanowire axis. Themeasured $r_{\text{MR}}$ and $r_{\text{MTP}}$ data include contributions from GMR, AMR, and magnon-related magnetoresistance. Below the saturation field, the GMR and AMR effects dominate, while above the saturation field, the magnon-related magnetoresistance dominates the magnetotransport[65]. Nernst effects of the contact structure and time-dependent changes due to a continuous temperature rise of the microstructure were determined from the slopes above saturation and were corrected, which resulted in symmetric hysteresis loops. The correction of these deviations, typically below 0.06 $\mu$V K$^{-1}$ at 1 T and RT, are important for the following comparisons of $S$ and $\sigma$.

## IV. RESULTS

### A. Magnetoresistance

The resistivities of the multilayered nanowires at RT vary between 28 and 50 $\mu\Omega$ cm, as shown in Fig. 4(a), with no clear dependence on the Cu layer thickness in contrast to the literature[66,67]. It seems that the zero-field resistivity is dominated by nonsystematic changes of the impurity concentration, the crystallinity, and lattice defects, but not the spin-dependent scattering, which should depend on the Cu layer thickness. The residual resistivity values obtained from the temperature behavior can be expected to be between 16 and 35 $\mu\Omega$ cm. These values are a factor of 2–5 higher than measured for electrochemically deposited Co-Ni alloy nanowires[9] and thin films[68], which seems reasonable by considering the additional scattering due to Cu impurities, interfaces, and spin-dependent scattering. Errors in the diameter measurement have to be considered and might lead to deviations up to 15% of the resistivity. Lenczowski *et al.*[66] investigated the current-in-plane (CIP) resistivity of electrochemically deposited Co/Cu thin films and found a decrease from 15 $\mu\Omega$ cm at 1-nm Cu layer thickness to 5 $\mu\Omega$ cm at 5 nm. The Cu layer thickness has a major influence on the GMR effect, which reaches a maximum for electrodeposited nanowires (CPP geometry) between 3 and 5 nm[35,42,69,70]. Bakonyi and Peter published a comprehensive review on the progress and difficulties regarding GMR multilayered film depositions[71]. The magnitude of the GMR effect is independent of the direction of the applied magnetic field. On the other hand, in the case of the bulk AMR effect, the resistivity increases for a magnetic field parallel to the measuring current and decreases if the current and magnetic field are perpendicularly aligned. Therefore, the GMR and AMR contributions can be distinguished by comparing the $r_{MR}$ values above the saturation field in parallel and perpendicular directions to the nanowire axis, as shown in Fig. 2. In Table II the total perpendicular $r_{MR\perp}$, parallel $r_{MR\parallel}$, and the difference of the values in both directions, the AMR effect $r_{AMR}$, are given. According to Liu *et al.*[72], three regimes can be distinguished in electrochemically deposited multilayers: continuous bilayers, pinholes in the nonmagnetic layer, and pinholes in the magnetic layer. At continuous Cu layers, the GMR effect typically decreases as the Cu layer thickness increases. Below a certain spacer thickness, pinholes form in the Cu layer and a direct exchange coupling leads to a parallel alignment of the magnetic layers in zero field. The 0.8-, 1.4-, and 5.2-nm Cu samples indicate a pinhole-dominated behavior with the increasing GMR effect with increasing spacer thickness. The 3.5-nm Cu nanowire has thinner magnetic segments and, as a result, a higher GMR effect. At thinner spacer thicknesses, the GMR effect vanishes and only the AMR effect remains (0.2- and 0.9-nm Cu samples). In the third case (not observed) the so-called superparamagnetic magnetoresistance occurs, due to magnetic islands embedded in the nonmagnetic matrix that act like magnetic nanoparticles. Our perpendicular magnetoresistance values above the saturation fields at RT are comprised between −3.3% and −15.5%see Fig. 4(b)], while the highest value for electrodeposited Co-Ni/Cu multilayered nanowires in the literature is −35% (derived from the $r_{MR,inf}$ value of −55% given in Ref. 32). The decrease of the general MR magnitudes between the two 3.5-nm Cu measurements 21 months apart can be ascribed to aging of the nanowires that are stored at RT in ethanol. Subsequent measurements three and seven months later show a further decrease of both the $r_{MR}$ value and the electrical conductivity.

## B. Magnetothermopower

The thermopower or Seebeck coefficient of a multilayered nanowire can be described as a series of Co-Ni and Cu segments, which results in a relation known from the so-called Nordheim-Gorter rule[52,73,74]. Using the compositional and geometrical data given in Table I and the thermopower values of Co-Ni alloy nanowires reported elsewhere[9], the literature bulk values of $S_{Cu}$[75], and bulk resistivities[16], the overall thermopower at RT is roughly estimated. Two effects increase the estimated thermopower with decreasing Cu layer thickness from −17 to −25 $\mu$V K$^{-1}$: These are the increasing ratio of Co-Ni to Cu and, more importantly, the compositional change of the magnetic layer (the latter resulting in a change of $S_{Co-Ni}$ from −18 to −24 $\mu$V K$^{-1}$). The measured thermopowers at RT vary between −15 and −24.5 $\mu$V K$^{-1}$ and fit well to the calculation. These thermopower RT values are given in Table II; the values of some samples are extrapolated from the trend at low temperatures, due to errors in the determination of $\Delta T$ above 150 K. The Seebeck coefficients in Fig. 4(c), interpolated at 25-K steps for clarity, show a monotonic increase with the temperature, as expected for most metals due to a single type of charge carrier and by the dominating diffusive thermopower[52].

As explained above, the thermopower for Co/Cu and Ni/Cu multilayers is estimated from the literature values for bulk in reference to Pt[16,21,76–78]. Since the composition has a stronger influence than the layer thicknesses, the Co-Ni to Cu ratio is set to 5:1. It seems that the measured values of the 0.9-nm Cu sample are shifted upward, while the values of the 3.5-nm Cu sample are shifted downward in comparison with the estimation. The phonon-drag peak around 75 K in the estimated curves is due to the bulk Pt values. This peak is not visible in the nanowire measurements as expected due to the nanostructuring of the Pt contacts. Above 150 K, some measurements show irreversible deviations of $\Delta T$. Wherever possible, the faulty data are replaced by thermometer data of a comparable sample and scaled to fit the low-temperature region. At each temperature step, the magnetic field dependence of the thermopower is measured. The relative change due to the applied magnetic field (the $r_{MTEP}$ value) is shown in Fig. 4(d). The RT values of the $r_{MTEP}$ are 2.7% for 0.2-nm Cu, 4.2% for 0.8-nm Cu, 3.5% for 0.9-nm Cu, 14.7% for 1.4-nm Cu, 29.1% for 3.5-nm Cu, and 14.5% for 5.2-nm Cu. These values are similar to the available literature values for Co/Cu thin films and nanowires, which range between 5% and 32%[1,2,44,73,74,79,80].

The measured $r_{MTEP}$ values are higher than the $r_{MR}$ values, which is in agreement with most nanowire and CPP thin-film literature[1,44,74,79,80]. On the contrary, two publications on CIP thin-film measurements show the opposite behavior with higher $r_{MR}$ than the $r_{MTEP}$[2,73]. Kobayashi *et al.*[81] studied the difference between CIP and CPP measurements on the same sample and found no systematic difference between the measurement directions. (More precisely, a current-atangle- to-plane contribution is measured that involves a CPP contribution.) Therefore, this variation in the literature data might not be due to the alignment of the current with respect to the multilayer planes, but rather due to the contact materials used and measurement setup. The $r_{MTEP}$ [Eq. (2)] depends on the thermopower of the electrical contact structure. Because Co/Cu and Co-Ni/Cu multilayers have negative Seebeck coefficients, positive/negative absolute Seebeck coefficients of the electrical contacts $S$abs contact will lead to decreased/increased $r_{MTEP}$ magnitudes compared to the $r_{MR}$ values (Fe, Au, and Cu are positive and Pt is negative at RT). In publications that specify the contact material, the decreased/increased $r_{MTEP}$ magnitudes seem to correlate with positive/negative $S$abs contact. Shi *et al.*[2] used Fe as contact material and measured decreased $r_{MTEP}$ values. Gravier *et al.*[1,44,74,79] state to have measured systematically too low Seebeck coefficients (despite statingAu as the contact material), which nevertheless explains the repeatedly observed increased $r_{MTEP}$ value[82]. Previous results on Co-Ni alloy nanowires with Pt and Au contacts behaved accordingly[9].

Our measured $r_{MTEP}$ at temperatures above 200 K is similar in absolute magnitude to the $r_{MR}$, but at lower temperatures, the samples can be arranged into two groups. The $r_{MTEP}$ of the 0.8- and 3.5-nm Cu samples continuously increases with decreasing temperature, while the rest of the samples reach a maximum around 180 K. In the case of the 3.5-nm Cu nanowire, both behaviors occur; therefore, the deviation is not due to the nanowires themselves. The explanation is that the $r_{MTEP}$ includes the thermopower of the electrical contact structure. Due to inconsistencies of the sputtering setup, the electrical contact structure of the 0.8- and 3.5-nm Cu samples incorporated Cr, which is known to diminish the phonon drag, e.g., in Au[83]. As shown by Huebener[24], the absolute thermopower of Pt crosses zero around 180 K and reaches a maximum around 70 K due to the phonon drag. Although no phonon-drag-related peak is observed in the thermopower measurements, the Pt contacts still lead to a maximum of the $r_{MTEP}$ at 180 K by decreasing/increasing the measured thermopower at lower/higher temperatures. In other words, the thermopower of the Pt contacts still shows the typical zero crossing between 150 and 200 K. By incorporating the Cr impurities, this zero crossing seems to be suppressed and the thermopower of Pt-Cr contact adds an approximately constant increase to the $r_{MTEP}$ value.

The absolute change of the Seebeck coefficient due to the magnetic field $\Delta S$ is between 0.66 and 4.4 $\mu$V K$^{-1}$. The $\Delta S$ is independent of the contact material, therefore, it is a useful property to compare the magnitudes of the effects of different materials. The highest $\Delta S$ values of around 8 $\mu$V K$^{-1}$ at RT are measured by Shi et al.[2] and Nishimura et al.[73] on sputtered Co/Cu thin films with a nonmagnetic layer thickness of 1 nm. These thin films showed CIP GMReffects of about 50%.

## V. CORRELATION BETWEEN THE THERMOPOWER AND THE RESISTANCE

The $S(H)$ and the corresponding $R(H)$ curves displayed in Fig. 3(b) indicate a typical linear relation between $S$ and $R^{-1}$ with a certain deviation above the saturation. It is important for the analysis that the conditions (e.g., temperature gradient and average temperature) during thermopower and resistivity measurement are identical; otherwise the dependences are not comparable. The linear relation between $S$ and $R^{-1}_{res}$, with $R_{res}$ being the residual resistance, was first found by Nordheim and Gorter[57] and was described more comprehensively by Gold et al.[56] with the impurity concentration as an implicit variable. Conover et al.[3] then predicted equal $r_{MTP}$ and $r_{MR}$ magnitudes and attempted to verify this experimentally. In the present work, the magnetic field is varied between ±3 T as an implicit variable and a linear relation between $S(H)$ and $R(H)^{-1}$ is found at each temperature, as shown in Fig. 5(a). This linear relationship in combination with the Mott formulasee Eq. (1)] indicates a magnetic-field-independent $d\rho/dE$ at the Fermi energy[1–14]. By fitting $S$ versus the conductance scaled by the average temperature of the nanowire ($\bar{T} R^{-1}$), the temperature-dependent energy derivative of the resistivity can be extracted from the slope and $S$offset can be extracted from the offset, as shown in Figs. 5(b) and 5(c).

A temperature-dependent increase of the slope has been published on Co/Cu multilayers by Baily et al.[80] and by Shi et al.[2,6,10], on Cu/Co/Cu/Ni-Fe multilayers by Kobayashi et al.[81], and on Fe-Ag granular alloys by Sakurai et al.[5]. Figure 5(c) shows the temperature-dependent offset $S_{\text{offset}}$ of the linear fits on the data shown in Fig. 5(a). The measured Seebeck coefficient is in reference to the contact material, while the resistance measurement gives the resistance of the nanowire. Any magnetic field dependence of the measured Seebeck coefficient $S$measured should be caused by the nanowire. In the following $S_{\text{NW}}$ and $S_{\text{contact}}$ refer to absolute thermopowers. Due to a magnetic-field-independent $d\rho/dE$ at the Fermi energy as discussed earlier, the Mott formula (1) predicts that the magnetic field dependence of the Seebeck coefficient is proportional to the nanowire conductivity $\rho_{\text{NW}}(H)^{-1}$ at any given temperature. This can be summarized in the following two formulas for the measured Seebeck coefficients:

$$S_{\text{measured}}(H) = S_{\text{NW}}(H) - S_{\text{contact}} \qquad (2)$$

$$S_{\text{measured}}(H) = -cT \left( \frac{d\rho}{dE} \right)_{E=E_F} \rho_{\text{NW}}(H)^{-1} - S_{\text{offset}} \qquad (3)$$

In general, $S_{\text{offset}}$ can arise from the following thermopower contributions: (i) thermopower of the electrical contacts ($S_{\text{contact}}$), (ii) nondiffusive thermopower of the sample (drag effects), and (iii) a magnetic-field-dependent energy derivative of the resistivity. The linear temperature behavior of the Seebeck coefficient of the investigated samples suggests a dominating diffusive behavior. In polycrystalline nanostructured samples of Co, Ni, and Cu, a nondiffusive thermopower can be most likely excluded, as discussed previously. Therefore, point (ii) can be neglected, but the thermopower contribution below 100 K should be carefully treated in general due to the high uncertainties and a wide range of possible effects. In the observed magnetic field range, the energy derivative of the resistivity is magnetic field independent as stated in the literature several times. Therefore, point (iii) is carefully rejected, leaving only point (i). For this material system, it follows that $S$contact $=S$offset and, as already predicted for Fe-Cr by Conover et al.[3],

$$r_{\text{MTP}} = (S(H) - S_0)/S_{\text{NW},0} = -(R(H) - R_0)/R(H) = -r_{\text{MR,inf}}. \qquad (6)$$

At the same time, the $r_{\text{MTP,inf}}$ is equal to $-r_{\text{MR}}$. By comparing the $r_{\text{MTP}}$ to $r_{\text{MR,inf}}$ or $S_{\text{contact}}$ to $S_{\text{offset}}$, a quantitative statement about the previous assumptions can be made. Figure 5 shows $S_{\text{offset}}$ and the absolute bulk Pt literature values, which fit qualitatively. The temperature-dependent literature values for absolute $S_{\text{Pt}}$ are a combination of the data by Roberts[21] (above

270 K) and Moore and Graves[77] (below 270 K), representing the most reliable literature data in each temperature region. Deviations are expected due to two effects: Size effects should reduce the phonon-drag peak[24] and impurities of the materials used in measurement setups can be expected to cause deviations from the pure bulk literature value[83]. The surface is likely contaminated by the necessary ac sputter cleaning process of 15-min duration. The Ti adhesion layer sputtered prior to the Pt deposition leads to a parallel circuit[84] of the Ti and the Pt layers and a deviation of about 0.5 $\mu$V K$^{-1}$ at RT[21,77,85–87]. The offsets of the three samples deviate from each other. Since all three electrical contact structures show very similar heating and resistance behavior, the deviations are unexpected and might be a sign for nondiffusive thermopower contributions of the nanowires. This seems questionable close to RT and is in conflict with the previous discussion on point (ii). Independently of the origin of the deviation between the samples, the calculation of the absolute thermopower leads to very good results, as shown in the following.

In Figs. 6(a) and 6(b), the temperature behavior of the absolute thermopower of three nanowires is calculated using the bulk Pt literature value [Fig. 6(a)] and $S_{contact} = S_{offset}$ [Fig. 6(b)]. Theoretical values of the absolute thermopower for Co/Cu and Ni/Cu multilayers with 5:1 layer thicknesses are estimated by using the absolute bulk literature values[16,21,76–78]. The correction by the absolute bulk literature value of Pt shifts each curve by a fixed value and changes the curvature in opposite direction, which suggests a positive phonon-drag or magnon-drag contribution similar to the results of Farrell and Greig[88] for bulk Ni. In general, bulk Co, Ni, and Cu metals show a significant phonon-drag contribution at 70 K, which is decreased in bulk Ni by adding Co impurities as Farrell and Greig showed[88]. In nanocrystalline metals, phonon transport is restricted and the phonon-electron scattering probability is thus reduced[89–91]. Hence, the phonon-drag thermopower in electrochemically deposited materials is typically negligible[9,27].

Overall, the deviations in Fig. 6(a) from the diffusive behavior are unreasonable and almost certainly are due to an artifact of the correction by inappropriate bulk values. The individual correction by $S_{offset}$ for each sample leads to the curves shown in Fig. 6(b). The curvatures are almost completely removed and the thermopower shows the expected linear temperature behavior without an offset at absolute zero. Comparing Figs. 6(a) and 6(b) leads to the conclusion that the correction by $S_{offset}$ is more appropriate for this material system.

The $r_{MTEP}$ values are corrected by $S$offset to obtain the $r_{MTP}$ values, which are compared to the $r_{MR,inf}$ values. The $r_{MTP}$ and $r_{MR,inf}$ curves shown in Fig. 7 have to match according to Eq. (6) and any deviations arise from variation from the linear fits in Fig. 7. The uncorrected $r_{MTEP}$ curves are added to the figure as lines, illustrating the significance of the correction. Nonmonotonic deviations as shown by the $r_{MTEP}$ curves are commonly attributed to drag effects of the sample[12,89,92]. Although these effects can dominate depending on the sample properties, the influence of measurement artifacts due to the contact material should be carefully considered. For instance, the $r_{MTEP}$ at RT decreased by up to 3% due to the influence of the electrical contacts, which explains why published $r_{MTEP}$ values deviate from the $r_{MR}$ values as discussed in the previous section. In the case of samples with dominating drag effects, this model can be used to quantify the deviations from the Mott formula by minimizing the contribution of the electrical contacts.

Several other coherent conclusions follow from this line of thought. According to Eq. (6), a finite $r_{MR}$ value and vanishing absolute $S$ result in a vanishing change $\Delta S$ of $S$ due to the magnetic field. In addition, a sign change of the absolute $S$ induces a sign change of $\Delta S$, which is exactly the result of one of the first $r_{MTEP}$ measurements by Piraux et al.[51] on Fe/Cr multilayers. The sign of $S$ is given by the charge of the carriers and the energy derivative of the resistivity. Therefore, in metals the sign of the energy derivative of the resistivity determines the sign of $S$ and $\Delta S$. In contrast, the $r_{MTEP}$ can have either sign or value due to various possible contact offsets, which is in agreement with experimental results and density of states evaluations by Tsymbal et al.[93] (The quantity $r_{MTEP}$ is used with the meaning of the quantity $r_{MTP}$ in Ref. 93.)

## VI. CONCLUSIONS

Current-perpendicular-to-plane MTP and GMR measurements on single Co-Ni/Cu multilayered nanowires are presented with varying thickness of the Cu spacer. The thermopower values of electrochemically deposited multilayered nanowires are measured to be −15 to −24.5 $\mu$V K$^{-1}$ at RT, which convincingly agrees with estimated values between −17 and −25 $\mu$V K$^{-1}$. Magnetoresistance measurements in parallel and perpendicular magnetic fields show that the thinnest Cu spacers are not continuous and these samples show negligible GMR effects. For samples with thicker Cu layers, the GMR effects are between −9% and −25% at RT. A linear relationship between the magnetic-field-dependent Seebeck coefficient $S$ and the electrical conductivity $\sigma$, with the magnetic field as an implicit variable, is found, as expected from the Mott formula, which describes the diffusive thermopower contribution. Disregarding nondiffusive thermopower contributions, a simple model is proposed to separate the absolute thermopower of the sample from the magnetic-field-independent thermopower of the contact material, without relying on the literature values of the latter. The temperature dependence of the thermopower offset agrees qualitatively with the literature values of the absolute Seebeck coefficient of the contact material. The absolute thermopower, the $r_{\mathrm{MTP}}$ values, and the energy derivative of the resistivity are calculated as a function of temperature. In accordance with the model, equal magnitudes of $r_{\mathrm{MR,inf}}$ and $r_{\mathrm{MTP}}$ values are the consequence. Although open questions remain, the methods presented provide a powerful tool to quantify and separate the different thermovoltage contributions.

## VII. ACKNOWLEDGEMENTS

The authors gratefully acknowledge financial support via the German Academic Exchange Service, Spanish MINECO under research Projects No. MAT2010-20798-C05-04 and No. MAT2013-48054-C2-2-R, the German Priority Program SPP 1536 "SpinCAT" funded by the Deutsche Forschungsgemeinschaft (DFG), and the excellence clusters "The Hamburg Centre for Ultrafast Imaging" funded by the DFG and "Nanospintronics" funded by the State of Hamburg. Scientific support from the University of Oviedo SCTs is also recognized.Work carried out in Budapestwas supported by the Hungarian Scientific Research Fund under Grant No. OTKA K104696.

TABLE I. Bilayer lengths $l_{bilayer}$ and magnetic $l_{Co-Ni}$ and nonmagnetic $l_{Cu}$ layer thicknesses of the investigated nanowires. The atomic ratios of Co:Ni:Cu and Co:Ni were obtained from the overall nanowire compositions measured by the transmission electron microscope (data given after a correction for the Cu grid background) or the scanning electron microscope (these data are marked with an asterisk). The bilayer thicknesses were determined as described in the text. The two thinnest $l_{Cu}$ values are roughly approximated.

| $l_{Cu}$ sample | $l_{bilayer}$ (nm) | $l_{Co-Ni}$ (nm) | Co:Ni:Cu | Co:Ni |
|---|---|---|---|---|
| 0.2-nm Cu | not measured | | not measured | not measured |
| 0.8-nm Cu | not measured | | not measured | not measured |
| 0.9-nm Cu | 17.3 ± 1.3 | 16.4 | 32:65:3 | 33:66 |
| 1.4-nm Cu | 17.5 ± 1.5 | 16.1 | 47:47:6 | 50:50 |
| 3.5-nm Cu | 8.7 ± 1 | 5.2 | 13:18:69 | 42:58 |
| | 9.2* | 5.4* | 14:14:72* | 50:50* |
| 5.2-nm Cu | 22.6 ± 1.1 | 17.4 | 21:50:29 | 30:70 |

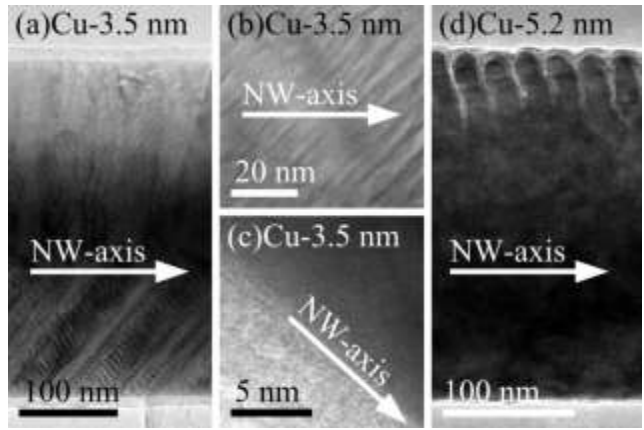

FIG. 1. (a) Transmission electron microscope image of a single nanowire (NW) of the 3.5-nm Cu sample with an average bilayer thickness of 8.7 nm. The fringes at the bottom are not related to the bilayered structure, but are rather artifacts that are associated with electron scattering at twin boundaries in the material because of the large nanowire diameter. (b) Magnification of the bilayers and (c) high-resolution transmission electron microscope image of the edge of the same nanowire and the $SiO_2$ shell in the bottom left. (d) Transmission electronmicroscope image of the 5.2-nm Cu sample.

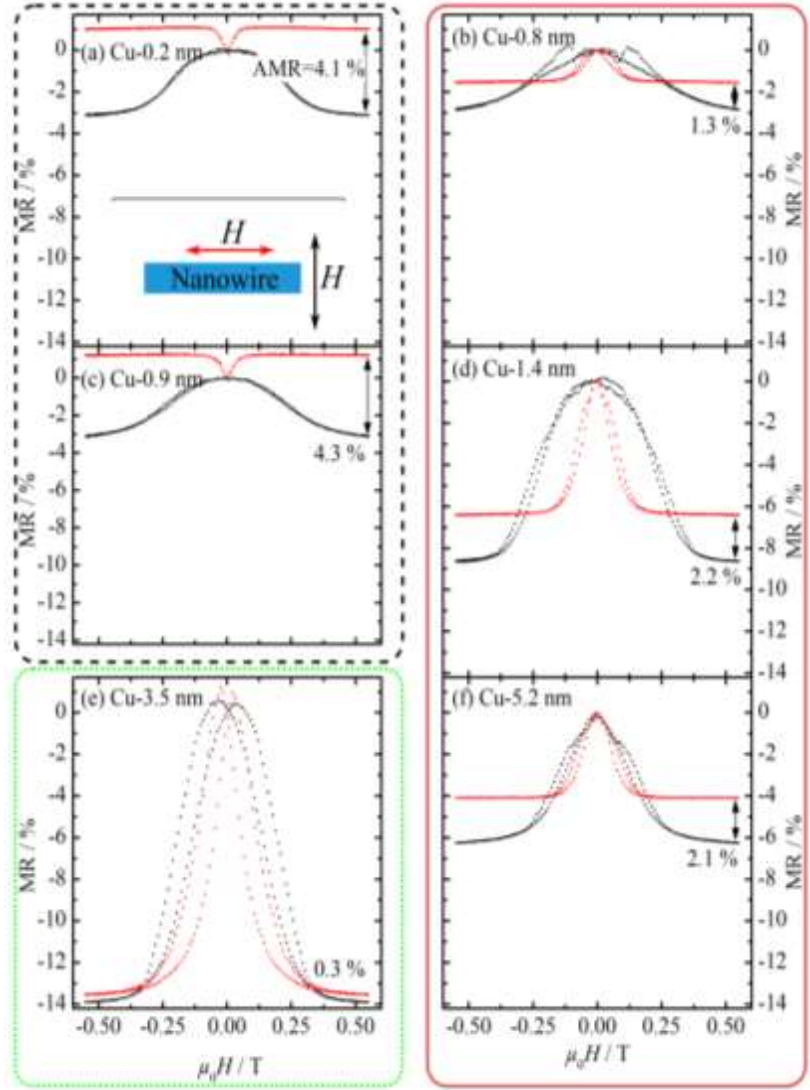

FIG. 2. Plot of the $r_{MR}(H)$ curves of the Co-Ni/Cu multilayered nanowires in parallel and perpendicular directions of the magnetic field with respect to the nanowire axis (electrical current direction) at RT for (a) 0.2-nm Cu, (b) 0.8-nm Cu, (c) 0.9-nm Cu, (d) 1.4-nm Cu, (e) 3.5-nm Cu, and (f) 5.2-nm Cu samples. The samples within the dashed black border show AMR-dominated behavior due to pinholes in the nonmagnetic layers. The samples within the red border show significant GMR effects due to continuous bilayers and the sample within the dotted green border shows a dominating GMR effect. The saturation fields are not reached for all of the samples and the actual saturation values are slightly higher.

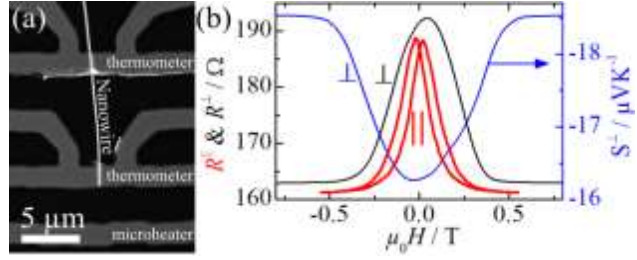

FIG. 3. (a) Scanning electron microscope image of a nanowire and the electrical contact structure. (b) Parallel (probe station setup) and perpendicular (cryostat) resistance measurements at RT and a Seebeck coefficient measurement (cryostat) in amagnetic field for the 3.5-nm Cu sample.

TABLE II. Room temperature values of resistivities, $r_{MR\perp}$, $r_{MR\parallel}$, $r_{AMR} = r_{MR\parallel} - r_{MR\perp}$, thermopower $S$, and magneto-thermoelectric power ($r_{MTEP}$) at RT are given. The $r_{MR\parallel}$ and $r_{MR\perp}$ data were obtained in a magnetic field of 0.55 T in parallel and perpendicular direction to the nanowire axis, respectively, as shown in FIG. 2. The saturation fields are not reached for all samples and the actual values are slightly higher. The MTEP data were obtained in perpendicular magnetic field up to 3 T.

| Sample-$l_{Cu}$ | $\rho$ ($\mu\Omega$cm) | $r_{MR\perp}$ (%) | $r_{MR\parallel}$ (%) | $r_{AMR}$ (%) | $S$ ($\mu$VK$^{-1}$) | $r_{MTEP}$ (%) |
|---|---|---|---|---|---|---|
| 0.2-nm Cu | 30.7 | -3.1 | +1.0 | 4.1 | -24.5 | 2.7 |
| 0.8-nm Cu | 36.8 | -2.9 | -1.6 | 1.3 | -18.5 | 4.1 |
| 0.9-nm Cu | 33.2 | -3.1 | +1.2 | 4.3 | -22.2 | 3.6 |
| 1.4-nm Cu | 50.8 | -8.6 | -6.4 | 2.2 | -15.7 | 14.8 |
| 3.5-nm Cu | 44.9 | -13.9 | -13.5 | 0.3 | -15 | 29.0 |
| 5.2-nm Cu | 28.7 | -6.2 | -4.1 | 2.1 | -15.7 | 14.3 |

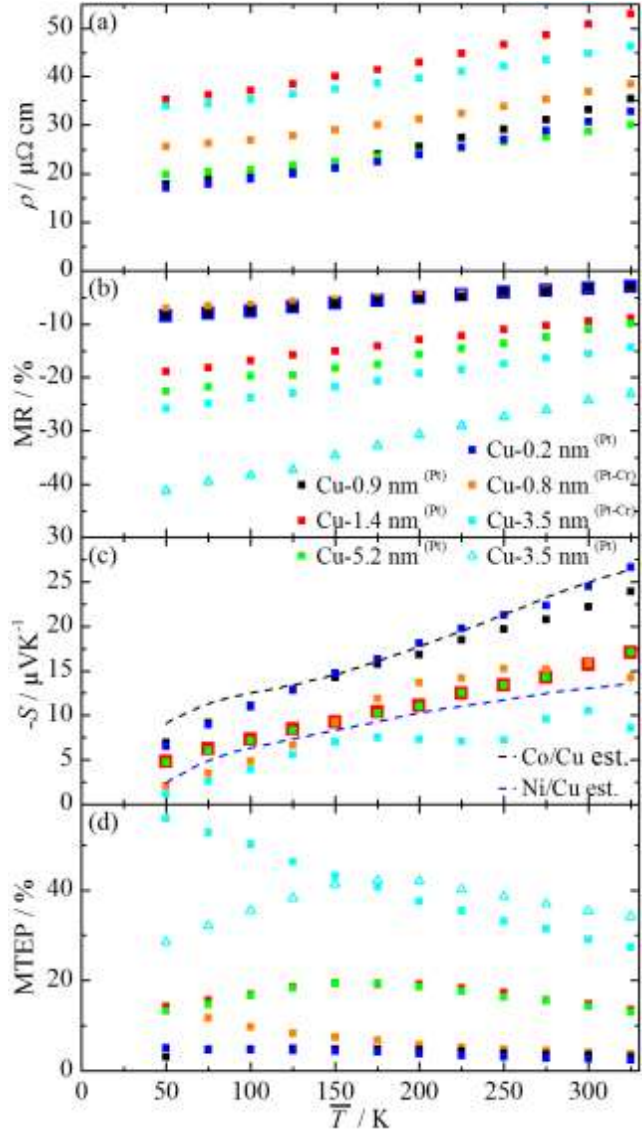

FIG. 4. Temperature dependence of the (a) zerofield resistivity, (b) $r_{MR}$, (c) thermopower, and (d) $r_{MTEP}$ of the multilayered nanowires. The $r_{MR}$ and $r_{MTEP}$ data were measured in perpendicular magnetic fields of 3 T. Also shown are (c) estimated thermopower values for Co/Cu and Ni/Cu multilayer with a layer thickness ratio of 5:1 using the literature bulk data for $S_{Co}$, $S_{Ni}$, and $S_{Cu}$. (b) and (d) Two 3.5-nm Cu samples were measured 21 months apart, which caused a reduction of the effect magnitudes.

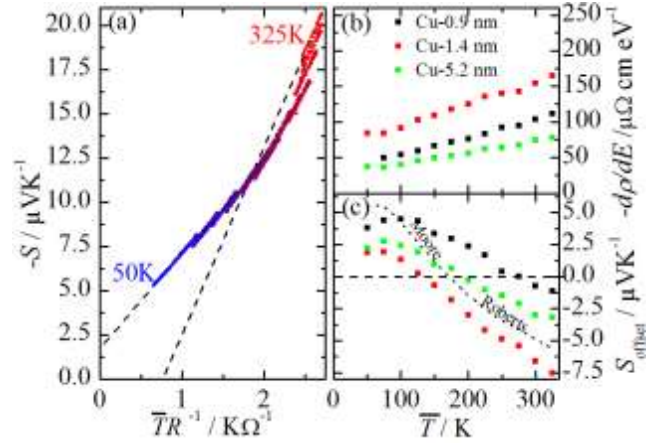

FIG. 5. (a) Seebeck coefficient versus average temperature times the conductance of the 1.4-nm Cu sample in 25-K steps from 50 to 325 K with the applied magnetic field as an implicit variable. For simplicity, only data for $U_{heater} = 5$ V are shown, which correspond to a $\Delta T$ of 3 K at 25 K to 2 K at 325 K. (b) Energy derivative of the resistivity at the Fermi energy derived from Eq. (5) against the temperature. (c) Offset from Eq. (5) and the absolute literature values of Pt[21,77].

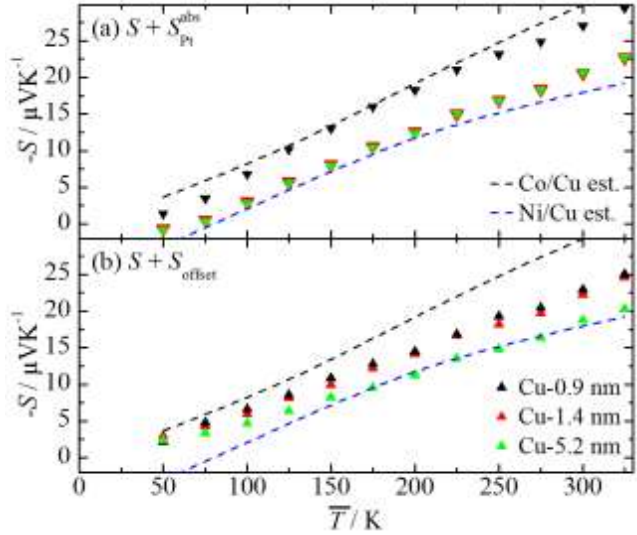

FIG. 6. (a) Absolute thermopower obtained by correcting by the literature values for $S_{Pt}$[21,77], $S + S^{abs}_{Pt}$, and (b) the absolute thermopower obtained by correcting by $S_{offset}$, $S + S_{offset}$. The dashed lines indicate the estimated absolute thermopower values for Co/Cu and Ni/Cu multilayers with 5:1 layer thicknesses[21,76,78].

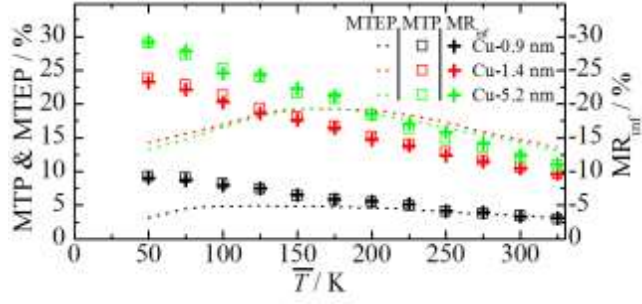

FIG. 7 Temperature dependence of $r_{\text{MR,inf}}$, $r_{\text{MTEP}}$, and $r_{\text{MTP}}$ values, which are corrected under the assumption of $S_{\text{contact}} = S_{\text{offset}}$. The expected relation between $r_{\text{MTP}}$ and $r_{\text{MR,inf}}$ according to Eq. (6) is observed.